\documentclass[conference,a4paper]{IEEEtran}
\IEEEoverridecommandlockouts

\usepackage{amsmath,amssymb,amsfonts}
\usepackage{graphicx}
\usepackage{cite}
\usepackage{url}
\usepackage{xcolor}
\usepackage{tikz}
\usetikzlibrary{positioning,fit,arrows.meta}
\usepackage{fancyhdr}
\begin{document}
	
\title{
RadioSim Agent: Combining Large Language Models and Deterministic EM Simulators for Interactive Radio Map Analysis
}

\author{\IEEEauthorblockN{
		Sajjad Hussain\IEEEauthorrefmark{1},   
		Conor Brennan\IEEEauthorrefmark{2}     
	}                                     
	\IEEEauthorblockA{\IEEEauthorrefmark{1}
		National University of Sciences and Technology (NUST), Islamabad, Pakistan, sajjad.hussain2@seecs.edu.pk}
	\IEEEauthorblockA{\IEEEauthorrefmark{2}
		Dublin City University, Dublin, Republic of Ireland, conor.brennan@dcu.ie} 
}

\maketitle

\begin{abstract}
Deterministic electromagnetic (EM) simulators provide accurate radio propagation modeling but often require expert configuration and lack interactive flexibility. We present \textbf{RadioSim Agent}, an agentic framework that integrates large language models (LLMs) with physics-based EM solvers and vision-enabled reasoning to enable interactive and explainable radio map generation. The framework encapsulates ray-tracing models as callable simulation tools, orchestrated by an LLM capable of interpreting natural language objectives, managing simulation workflows, and visually analyzing resulting radio maps. Demonstrations in urban UAV communication scenarios show that the agent autonomously selects appropriate propagation mechanisms, executes deterministic simulations, and provides semantic and visual summaries of pathloss behavior. The results indicate that RadioSim Agent provides multimodal interpretability and intuitive user interaction, paving the way for intelligent EM simulation assistants in next-generation wireless system design.
\end{abstract}


\section{Introduction}

Accurate radio propagation modeling is fundamental to the design, optimization, and analysis of emerging 6G wireless communication systems \cite{wang20206g}. Deterministic electromagnetic (EM) simulators, such as ray-tracing-based solvers, can provide accurate predictions of signal propagation. However, they often demand expert configuration, follow rigid workflows, and offer limited semantic interpretability and automation.

Recent advances in \textit{agentic artificial intelligence (AI)}, which enhance large language models (LLMs) with capabilities such as external function calling (tools), strategic planning, memory, and self-evaluation, have opened new directions for autonomous reasoning and decision-making in wireless communication systems. For instance, Jiang \textit{et al.}~\cite{jiang2024large} introduced a multi-agent architecture for 6G communication management that coordinates retrieval, planning, and evaluation agents to execute network-level tasks. Similarly, Tong \textit{et al.}~\cite{tong2025wirelessagent} proposed \textit{WirelessAgent}, an LLM-driven system integrating perception, memory, and planning modules for intelligent resource allocation and network slicing. Extending this concept, the same group presented \textit{A-Core}~\cite{tong2025acore}, a collaborative agentic framework for 6G core networks that employs a Core-Network Large Model (CN-LM) and multiple generative agents for autonomous service creation. These studies highlight the promise of agentic AI for adaptive and self-organizing network intelligence, yet their focus remains primarily on system-level management rather than physical-layer modeling.

Parallel research has explored LLMs for data-driven wireless modeling. Liu \textit{et al.}~\cite{liu2025llm4wm} proposed \textit{LLM4WM}, a Mixture-of-Experts and Low-Rank Adaptation (MoE-LoRA) framework that fine-tunes pre-trained LLMs for channel estimation, beam management, and localization. By aligning wireless data with semantic representations, LLM4WM demonstrated strong multi-task generalization. Shi \textit{et al.}~\cite{shi2025geonr} introduced \textit{GeoNR-PSW}, which encodes ray-traced 5G channel fingerprints into pseudo-signal words, enabling few-shot localization via a frozen GPT-2 model with lightweight adapters. Although these methods connect linguistic reasoning with wireless data, they remain reliant on precomputed datasets and do not support autonomous EM simulation.

A closely related study by Quan \textit{et al.}~\cite{quan2025llmradio} proposed an LLM-agent-based framework for automating radio map generation and wireless planning. Their approach streamlines commercial EM tool workflows, improving usability and efficiency. However, it operates as a task automation layer without direct access to underlying EM computations, limiting transparency and reproducibility for research purposes.

Despite these advances, existing agentic and LLM-driven frameworks primarily operate at the network-management or data-analysis level. They optimize configurations or interpret measurements but do not \textit{simulate, analyze, or explain} radio propagation using first-principles EM computation. As a result, the potential of agentic AI to bridge natural-language understanding with deterministic physical modeling remains largely unexplored.

To address this gap, we introduce RadioSim Agent, an open-source, agentic framework that unifies LLM-based reasoning with deterministic EM solvers for interactive, multimodal, and explainable radio-map generation. Unlike prior works limited to data interpretation or network optimization, RadioSim integrates physics-driven modules for pathloss computation as callable tools within an agentic architecture. Beyond simulation, the agent also uses vision-enabled capabilities that allow it to query, interpret, and reason over generated pathloss heatmaps, providing semantic summaries, regional statistics, and propagation insights. Users can thus provide natural-language instructions to perform simulations, visualize EM fields, and interrogate results directly within a unified agentic environment.

The main contributions of this work are as follows:
\begin{itemize}
	\item We design an agentic EM simulation framework capable of generating pathloss and auxiliary feature maps (e.g., LOS mask, reflection maps, 3D distance fields) based on user-specified natural language instructions.
	\item We introduce multimodal reasoning capabilities, enabling the agent to interpret and explain radio maps via vision-language understanding.
	\item We demonstrate that the framework offers semantic interpretability, adaptive control, and interactive experimentation.
	\item The proposed agentic framework, including all simulation tools, reasoning modules, and workflow scripts, is openly available to the research community for reproducibility and extension.\footnote{The complete implementation is available at: \url{https://github.com/sajjadhussa1n/radio-sim-agent}}
\end{itemize}

Through these contributions, RadioSim Agent advances the frontier of LLM-driven scientific reasoning by bridging natural language queries, numerical EM simulation, and visual analysis. This establishes a foundation for open, explainable, and intelligent electromagnetic simulation environments for future 6G and beyond systems.

\section{Proposed Framework}

The proposed \textbf{RadioSim Agent} integrates deterministic EM simulation with agentic reasoning and visual understanding to enable interactive, explainable, and automated radio environment modeling. The framework allows users to describe simulation or analysis objectives in natural language, which are interpreted by an LLM that plans and executes EM-based propagation simulations and visual analyses through callable tool interfaces.

\begin{figure}[t]
	\centering
	\resizebox{0.95\linewidth}{!}{
		\begin{tikzpicture}[node distance=1.2cm, >=latex, auto, thick, align=center]
		
		\tikzstyle{module} = [rectangle, rounded corners, draw=black!80, fill=blue!10, thick,
		text width=4.5cm, minimum height=1.1cm, font=\footnotesize, align=center]
		\tikzstyle{arrow} = [->, line width=0.8pt, >=latex]
		\tikzstyle{databox} = [rectangle, draw=black!70, fill=green!10, thick,
		text width=3.8cm, minimum height=0.8cm, font=\footnotesize, align=center]
		\tikzstyle{input} = [rectangle, draw=black!50, fill=yellow!10, thick,
		text width=3.8cm, minimum height=0.8cm, font=\footnotesize, align=center]
		
		\node[input] (input) {User Prompt \\ \textit{(Natural Language Instruction)}};
		
		\node[module, below=1.1cm of input] (planner) {\textbf{Natural Language Planner (LLM)}\\
			Parses intent, extracts parameters, plans simulation or visual analysis actions, requests clarification if ambiguous.};
		
		\node[module, below=1.1cm of planner, xshift=-2.5cm] (simtools) {\textbf{Simulation Tool Library}\\
			Deterministic EM functions for LOS, REF, GREF, NLOS, and building losses; main callable: \texttt{simulate\_radio\_environment()}.};
		
		\node[module, below=1.1cm of planner, xshift=2.5cm] (visiontool) {\textbf{Vision Reasoning Tool}\\
			Interprets heatmaps, estimates pathloss distributions, and provides semantic summaries of visual data.};
		
		\node[module, below=3.0cm of planner, yshift=-0.3cm] (executor) {\textbf{Execution \& Output Module}\\
			Runs simulations, manages datasets, visualizations, metadata, and summarizes textual \& visual outputs.};
		
		\node[databox, below=1.1cm of executor] (output) {Results and Visualizations \\ \textit{Pathloss maps, datasets, visual summaries}};
		
		\draw[arrow] (input) -- (planner);
		\draw[arrow] (planner) -- (simtools);
		\draw[arrow] (planner) -- (visiontool);
		\draw[arrow] (simtools) -- (executor);
		\draw[arrow] (visiontool) -- (executor);
		\draw[arrow] (executor) -- (output);
		
		\node[rectangle, draw=black!60, dashed, inner sep=8pt, fit=(planner)(simtools)(visiontool)(executor),
		label={[xshift=3.0cm, yshift=0.1cm]\textbf{Reason–Act–Observe Cycle}}]{};
		
		\end{tikzpicture}
	}
	\caption{Architecture of the proposed RadioSim Agent integrating LLM reasoning with deterministic EM simulation and vision-based analysis for multimodal understanding.}
	\label{fig:radiosim_architecture}
\end{figure}
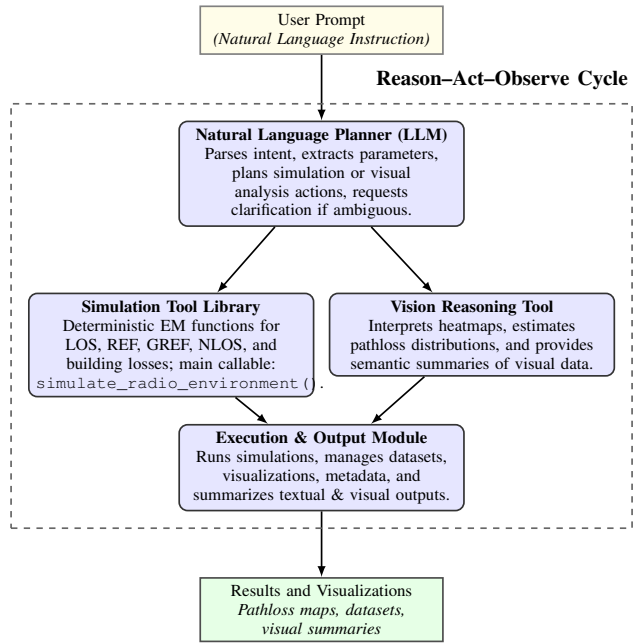

\subsection{Framework Overview}

As shown in Fig.~\ref{fig:radiosim_architecture}, the framework comprises four main components: 

\textbf{(1) Natural Language Planner:} 
The planner serves as the reasoning core of the RadioSim Agent and is implemented using the OpenAI GPT-4o-mini model. It interprets user instructions expressed in natural language and translates them into structured simulation or analysis tasks. Specifically, the planner extracts relevant parameters such as environment name, transmitter coordinates, receiver grid size, and selected propagation mechanisms. Leveraging its reasoning capability, the model decomposes complex objectives into sequential tool calls, determines dependencies among tasks, and orchestrates their execution. When user prompts are ambiguous or underspecified, the planner proactively issues clarification questions before proceeding ensuring robust intent understanding and reproducibility of results.

\textbf{(2) Simulation Tool Library:} 
The Simulation Tool Library provides deterministic EM modeling functions for computing key propagation mechanisms, including line-of-sight (LOS), wall reflections (REF), and ground reflections (GREF). The module supports five urban radio environments, namely Munich01, Munich02, London, Helsinki, and Manhattan, each represented using building vector data to capture realistic urban geometry.

The underlying ray-tracing engine computes direct and reflected propagation paths between the transmitter and a grid of receiver points, producing deterministic pathloss and associated feature maps \cite{HussainEucap20, HussainTAP22}. For receiver locations where no valid ray path is available, the module applies a 3GPP-based empirical pathloss model \cite{haneda20165g} to approximate non-line-of-sight (NLOS) attenuation, ensuring full spatial coverage. In addition, building entry loss (BEL) is computed following ITU-R P.2109 recommendations \cite{ITU-R-P2109-2} to account for signal penetration through walls.

All these computations are encapsulated within a Python-based callable tool, \texttt{simulate\_radio\_environment()}, which accepts user-specified parameters such as environment name, transmitter coordinates, grid resolution, and active propagation mechanisms. This modular design enables flexible configuration of physical models or selective inclusion of specific contributions thereby balancing accuracy, interpretability, and computational efficiency.

\textbf{(3) Vision Reasoning Tool:}
The Vision Reasoning Tool extends the agent’s analytical capability to visual modalities by integrating a vision-enabled large language model (LLM). In this work, we employ OpenAI GPT-4o-mini for multimodal reasoning on generated pathloss heatmaps. The tool interprets visual outputs by describing spatial signal patterns, estimating the approximate distribution of pathloss values, and identifying regions of degradation or obstruction. By combining semantic understanding with visual inference, this module enables post-simulation inspection, validation of propagation effects, and natural-language querying directly on visual data, thus bridging the gap between numerical output and intuitive interpretation.

\textbf{(4) Execution and Output Module:} 
The Execution and Output Module orchestrates the end-to-end simulation workflow, including task execution, metadata management, and result generation. It records all relevant simulation parameters, such as transmitter coordinates, environment identifiers, and propagation configurations to ensure traceability and reproducibility. The module also consolidates visual and numerical results into structured datasets for subsequent analysis. Through seamless integration with both the Simulation Tool Library and the Vision Reasoning Tool, it supports natural-language interaction over generated datasets and visual artifacts, enabling explainable and consistent post-simulation reasoning.

\subsection{Workflow}

The RadioSim Agent operates through a structured \textbf{Reason–Act–Observe} cycle.
In the \textit{Reason} phase, the LLM interprets the user’s intent expressed in natural language, extracts relevant parameters, and formulates an executable simulation or analysis plan.
During the \textit{Act} phase, the framework executes the required EM simulation functions to compute the selected propagation components using geometric and empirical models.
In the \textit{Observe} phase, the agent analyzes and summarizes the generated outputs. When visual interpretation is needed, the Vision Reasoning Tool is invoked to extract insights directly from the pathloss heatmaps, providing a multimodal summary that combines visual and textual reasoning. This unified workflow supports automated dataset generation, semantic post-simulation analysis, and interactive visual interpretation of radio propagation behavior.

\section{Experimental Setup and Results}

To demonstrate the reasoning and execution capabilities of the proposed \textit{RadioSim Agent}, we evaluate its ability to interpret natural language prompts, extract relevant simulation parameters, and autonomously invoke deterministic EM tools for radio-map generation. 

\subsection{Agent Execution Example}
We evaluated the agent’s ability to autonomously run radio propagation simulations using the following natural language prompt:

\begin{quote}
	\textit{“Simulate pathloss in the Munich01 scenario with a UAV at (100, 100, 15) over a 50×50 receiver grid considering all propagation mechanisms, and provide a concise technical summary of the resulting pathloss heatmap.”}
\end{quote}

Upon receiving the prompt, the agent parsed the request and determined which simulation functions to invoke. The execution trace is provided below:

\medskip

\noindent \textbf{Thought:}
\textit{“The user requests a simulation of the 'Munich01' environment with a UAV at (100,100,15), including all propagation components, and a summary of the pathloss heatmap.”}
\medskip

\noindent \textbf{Action:} 
\textit{“Call simulate\_radio\_environment with the specified parameters.”}
\medskip

\noindent \textbf{Action Input:}
\textit{“tx\_x = 100, tx\_y = 100, tx\_z = 15, location = 'munich01', nx = 50, ny = 50, LOS = True, REF = True, GREF = True, NLOS = True, BEL = True”}
\medskip

\noindent \textbf{Observation:} 
\textit{“Simulation completed successfully. The pathloss map was saved at data/pathloss.png”}.
\medskip

\noindent \textbf{Action:} 
\textit{“Call summarize\_pathloss\_image on the generated map.”}
\medskip

\noindent \textbf{Action Input:}
\textit{“image path = 'data/pathloss.png'”}
\medskip

\noindent \textbf{Observation:} 
\textit{“The pathloss heatmap shows values ranging from 110~dB (strong signal) to 170~dB (weak signal). Strong signals concentrate in the lower-left quadrant, while weaker signals dominate the upper-right quadrant. Gradients near walls and reflective surfaces indicate significant attenuation.”}
\medskip

\noindent \textbf{Final Answer:} 
\textit{“Simulation completed successfully. The pathloss heatmap shows strong signals in the lower-left quadrant and weaker signals in the upper-right quadrant, with notable gradients near reflective surfaces indicating areas of high attenuation.”}

\medskip

The simulation produced deterministic pathloss maps along with an auxiliary dataset text file containing key propagation features, including transmitter coordinates, receiver grid positions, line-of-sight (LOS) mask, azimuth angle ($\phi$), three-dimensional (3D) distance, reflection mask (indicating receiver points with valid reflection paths), building mask, and height map representing both building and receiver elevations and pathloss. The resulting dataset enables subsequent model training, validation, and detailed propagation analysis.

\begin{figure}[h]
	\centering
	\includegraphics[width=0.95\linewidth]{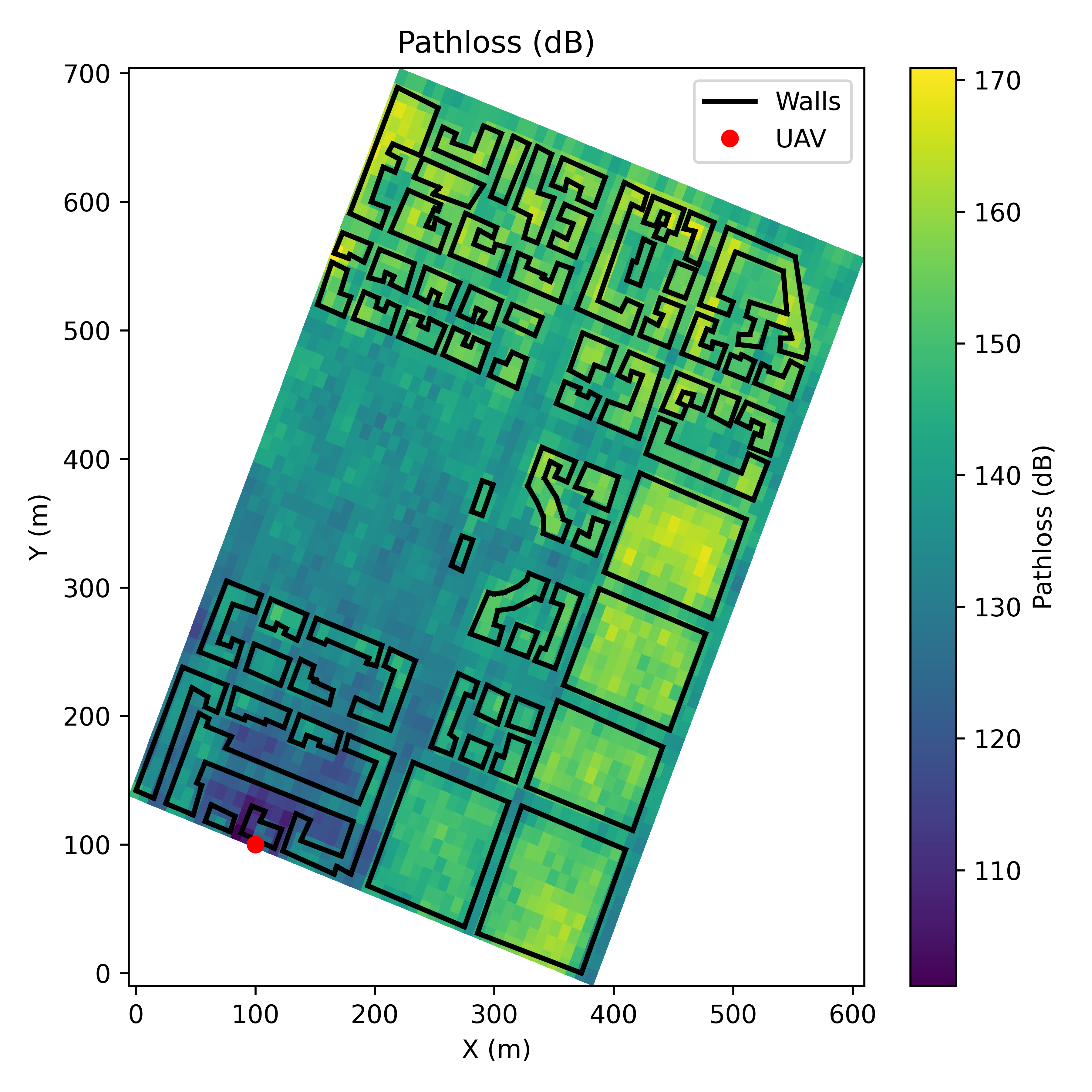}
	\caption{Pathloss heatmap generated by the agent for the Munich01 environment with a UAV at (100,100,15). Strong signals (lower pathloss) are observed in the lower-left quadrant, while weaker signals (higher pathloss) appear in the upper-right quadrant.}
	\label{fig:radio_maps}
\end{figure}

The generated pathloss heatmap for the simulated environment is presented in Fig.~\ref{fig:radio_maps}, showing clear spatial variations between strong and weak signal regions, in agreement with the agent’s summarized interpretation.

\section{Conclusion and Future Work}

This paper introduced the \textbf{RadioSim Agent}, an agentic simulation framework that unifies deterministic EM simulations with LLM reasoning for interactive and explainable radio propagation analysis. Unlike conventional solvers requiring expert configuration, the framework executes EM computations through natural-language control, autonomously invoking ray-based modules such as line-of-sight estimation, reflection modeling, and building-entry loss computation. The results demonstrate that RadioSim Agent can generate accurate and interpretable pathloss maps while providing coherent semantic explanations of propagation behavior. Future extensions will focus on integrating deep learning models to accelerate simulations, and enabling multi-agent collaboration for specialized propagation and optimization tasks. This work highlights the potential of LLM-driven EM simulation as a step toward intelligent, explainable modeling for next-generation wireless systems.

\bibliographystyle{IEEEtran}

\end{document}